\documentclass[a4paper,11pt]{article}
\usepackage{jcappub} 

\pdfoutput=1 

\usepackage{jcappub} 
\usepackage{pdflscape}
\usepackage[T1]{fontenc} 
\usepackage{graphicx}
\usepackage{epsfig}
\usepackage{rotate}
\usepackage{amsmath}
\usepackage{amssymb}
\usepackage{amsfonts}
\usepackage{multirow}
\usepackage{tablefootnote}
\usepackage{diagbox}
\usepackage{slashbox}
\usepackage{bm}
\usepackage{enumerate}
\usepackage{afterpage}
\usepackage{xcolor}
\usepackage{natbib, hyperref}
\usepackage{cancel}
\usepackage{amsmath,amssymb}

\def\be{\begin{equation}}
\def\ee{\end{equation}}
\def\bea{\begin{eqnarray}}
\def\eea{\end{eqnarray}}

\newcommand{\red}[1]{{\color{black}{#1}}}

\newcommand{\cc}[1]{{\color{black}{#1}}}
\newcommand{\pritha}[1]{{\color{black}{#1}}}
\allowdisplaybreaks

\title{ Visualising relativistic effects in redshift space distortions of large scale structure}

\author{Pritha Paul$^{1}$, Chris Clarkson$^{1,2}$}
\affiliation{ $^{1}$Department of Physics \& Astronomy, Queen Mary University of London, London E1 4NS, UK \\
$^2$Department of Physics \& Astronomy, University of the Western Cape, Cape Town 7535, South Africa
}

\emailAdd{p.paul@qmul.ac.uk}

\abstract{
Observing large scale structure in redshift space gives rise to the well known redshift space distortions whereby a spherical distribution of galaxies is distorted into an \red{ellipsoid} along the line of sight of the observer. This effect is important on linear scales  
and so can be thought of as a Newtonian correction to the density perturbation even though their physical origin is in the Doppler effect. On larger scales subtler aspects of the Doppler  and gravitational redshift effects give rise to further distortions in redshift space. These further contort objects beyond an \red{ellipsoidal} compression, into shapes with broken line-of-sight symmetry such as an egg- or bean-like shapes. In this paper, we aim to qualitatively picture how large over-dense regions, including clusters or superclusters, and under-dense regions, such as voids, undergoing infall or outflow respectively, become distorted in redshift space when higher-order relativistic effects are taken into account. 
This will contribute when analysing structure in real space such as stacking voids which are no longer radially symmetric when these effects are included.
}

\begin{document}
\maketitle
\flushbottom
\section{Introduction}

In redshift space, peculiar velocities along the line of sight distort the apparent positions of galaxies, leading to redshift space distortions (RSD), or the Kaiser effect \cite{1987MNRAS.227....1K}. On linear scales, this causes spherical over- or under-densities to appear squashed or stretched in ellipse-like shapes (for a comprehensive review see \cite{Hamilton_1998}).  However, on very large scales, additional distortions become relevant. These include asymmetric distortions along the line of sight, which go beyond the symmetric squashing described by the Kaiser effect. These relativistic contributions, mainly originating from the Doppler and gravitational redshift, are suppressed at least by a factor of $\mathcal{H}/k$ relative to the leading RSD terms (\cc{here, $\mathcal{H}$ is the conformal Hubble rate and $k$ is the comoving wavenumber}). Although these effects are subdominant on small to intermediate scales, they become increasingly important on large scales and cannot be neglected \cite{Yoo_2009,Yoo_2010,Bonvin_2011,Challinor_2011}. These projection effects were also extended to second order \cite{Bertacca_2014,Dio_2014,Yoo_2014} and more recently to third order \cite{Dio_2019,Paul_2024}. 

The relativistic corrections to the observed galaxy number density become particularly important in the case of a multi-tracer analysis of the $2$ point correlation function. Several groups have studied these corrections and shown that they give rise to odd multipoles in the cross-correlation between two galaxy populations \cite{Yoo_2009,McDonald_2009,Bonvin_2011,Challinor_2011}. In particular, \cite{Bonvin_2014} studied the correlation function between a bright and faint galaxy population, highlighting the asymmetry when exchanging their positions along the line of sight. This work provided a schematic representation of the physical effects involved, distinguishing between those that contribute symmetric and asymmetric terms. The Doppler effect and gravitational redshift in particular leave distinct imprints in the dipole of the observed signal. For higher-order statistics such as the bispectrum \cite{Clarkson_2019,PhysRevD.102.023530} and the trispectrum \cite{Paul:2024xff}, relativistic effects become relevant even within a single galaxy survey, particularly on large scales. 

To develop a more intuitive understanding of the nature of these effects beyond the simple linear Newtonian approximation, it is crucial to recognise that we observe galaxies in redshift space, not real space. This distortion is not an intrinsic feature of the real-space distribution, but rather a projection effect. While such distortions complicate the interpretation of galaxy clustering, they also encode valuable information about both Newtonian and relativistic contributions. The Newtonian approximation is sensitive to galaxy bias, whereas in the relativistic case, additional dependencies are introduced because of the evolution bias and magnification bias (and derivatives thereof). 
Importantly, the resulting odd multipoles are able to isolate these relativistic effects in large scale structure \cite{Saga:2020tqb,Breton_2018,Fonseca:2021gve}.

While the physical origin of these relativistic contributions has been studied at linear order, the resulting visual distortion of idealised objects in redshift space has not been considered in any detail. An exception is the work of \cite{Cai_2017}, who derived an expression for the observed redshift including relativistic corrections up to second order in velocity and applied it to the cluster-galaxy cross-correlation using N-body simulations. Their work showed that the relativistic second order terms introduce asymmetries and are sensitive to the non-spherical structure of halos. In this paper, we take a complementary approach to systematically investigate how these effects modify the over- and under-densities in redshift space up to second order in perturbation theory. Specifically, rather than consider a distribution of galaxies and systematically perturb each one to its redshift-adapted position, we look at the density field directly as derived in perturbation theory. This allows us to consider the variety of non-linear terms which appear and consider how structures get distorted from each of them in turn. We consider two idealised cases: a spherical cluster, modeled as an isothermal sphere, and a spherical void, described using a standard void density profile. We explicitly separate Newtonian and relativistic contributions in the observed galaxy number counts to investigate their effects in the above cases. We expect the Newtonian contributions to preserve symmetry along the line of sight, while relativistic terms introduce distinct asymmetries at the perturbative orders we consider. 

The outline of the paper is as follows: In Section~\ref{sec2}, we provide a brief overview of the galaxy number counts at first and second order and introduce the general formalism for relativistic RSD. Section~\ref{sec3} describes the setup for the line of sight vector and the derivation of the peculiar velocity component along it. We also present numerical results for the two profiles mentioned above: the isothermal sphere and the spherical void.


\section{\cc{The number count fluctuation in redshift space}} \label{sec2}

The fluctuations in the galaxy number counts $\Delta$ can be expressed as a function of the observable redshift $z$ and the observed direction $\boldsymbol{n}$, 
\bea \label{numcounts}
\Delta(z,\boldsymbol{n}) = \frac{N(z,\boldsymbol{n}) - \langle N \rangle (z) }{\langle N \rangle (z)}\,, 
\eea
where $\langle..\rangle$ is the average over observed direction $\boldsymbol{n}$ at fixed observed redshift $z$ and $N(z,\boldsymbol{n})$ is the number of galaxies. We \cc{expand \eqref{numcounts} perturbatively up to order $\mathcal{H}/k$ and we then we will proceed to investigate}  the visualisation of the different terms \cc{which originate from} \eqref{numcounts}. 

\subsection{First Order}
The redshift space distortions, including the relativistic corrections, to the galaxy number counts have been calculated by several groups \cite{Yoo_2009,Yoo_2010,Bonvin_2011,Challinor_2011}. \cc{The dominant local terms are }
\begin{align} 
        \Delta(z,\boldsymbol{{n}}) &= b\delta - \frac{1}{\mathcal{H}} \partial_{\chi} v_{||} + \frac{1}{\mathcal{H}}v'_{||} + \left( 1 - \frac{\mathcal{H}'}{\mathcal{H}^2} - \frac{2}{\chi \mathcal{H}} \right) v_{||} + \frac{1}{\mathcal{H}} \partial_\chi \psi\,.\label{delta_full}
\end{align}
The dash denotes a partial derivative with respect to conformal time. Here, $v_{||} = \boldsymbol{v} \cdot \boldsymbol{n}$, where $\boldsymbol{v}$ is the peculiar velocity of the galaxy. \cc{Consistently with our approximation we have ignored integrated effects and subdominant potential terms.} \pritha{It is also important to note that the relativistic contribution in (\ref{delta_full}) includes not only terms proportional to $v_{\parallel}$ but also contribution from the gradient of the gravitational potential, $\partial_\chi\psi$. While this gravitational potential term might initially seem separate from the velocity contributions, it is connected to them via the Euler equation. 
As a result, the gravitational redshift contribution is effectively of the same order as the velocity induced effects. The Euler equation takes the form 
\begin{align}
v'_{\parallel} + \mathcal{H}v_{\parallel} + \partial_\chi \psi = 0\,.
\end{align}
This relation allows $\partial_\chi \psi$ to be written in terms of $v_{||}$ and its conformal time derivative. When this is substituted in (\ref{delta_full}), the terms involving $v'_{\parallel}$ and $\partial_\chi \psi$ combine, resulting in a more compact form 
\begin{align} 
        \Delta(z,\boldsymbol{{n}}) &= b\delta - \frac{1}{\mathcal{H}} \partial_{\chi} v_{||} 
        - \Big(\frac{{\mathcal{H}'}}{\mathcal{H}^2} + \frac{2 - 5s}{\chi \mathcal{H}} + 5s -b_{\rm e} \Big) v_{||}\,, \label{fullnumc}
\end{align}
where additional contributions involving the magnification bias, $s$ and the evolution bias $b_e$ appear.} The galaxy distribution is related to the underlying dark matter through three bias parameters, $b$, the galaxy bias, $b_e$, defined as $b_e = \partial\mathrm{ln}(a^3\bar{n}_g)/\partial\mathrm{ln}a$, and $s$, given by $s = -(2/5)\partial \mathrm{ln}\bar{n}_g/ \partial\mathrm{ln}L$.
\cc{It it helpful to split these into Newtonian and relativistic contributions according to}  
\bea 
\label{Delta1N}
\Delta^{(1)}_N &=& b\delta - \mathcal{H}^{-1} \partial_{\chi} v_{||} \,,
\\
\label{Delta1R}
\Delta^{(1)}_R &=& \Big(\frac{{\mathcal{H}'}}{\mathcal{H}^2} + \frac{2 - 5s}{\chi \mathcal{H}} + 5s - b_e \Big) v_{||}\,.
\eea
The Newtonian part \cite{1987MNRAS.227....1K} \cc{contains the underlying}  fluctuations in the distribution of galaxies \cc{and} is accompanied by the redshift space distortion effect, which is given by the Kaiser term. The relativistic term describes the so-called Doppler effect \cite{Gaztanaga_2017,Giusarma:2017xmh,Lepori:2017twd,Breton_2018,Hall_2017,Yoo_2012,Bonvin_2016,Bonvin_2017}. This term is physically different from the standard Newtonian term as it is not symmetric along the line of sight and is also responsible for generating a dipole in the multi-tracer 2-point correlation function.
\subsection{Second Order}

At second order the galaxy number counts\cc{, including the relativistic corrections,} have  been derived by 
\cite{Bertacca_2014,Dio_2014,Desjacques_2018, Dio_2019, Clarkson_2019, Fuentes_2021} \cc{(see \cite{Bernardeau_2002, Nielsen_2017} for the Newtonian RSD derivation)}. \cc{The Newtonian and relativistic contributions relevant for us are given by:}
\bea
\label{Delta2N}
\Delta^{(2)}_N &=&\delta_g^{(2)} - \mathcal{H}^{-1} \partial_{\chi} v_{||}^{(2)} 
- \mathcal{H}^{-1} \partial_{\chi} \left( v_{||}  \delta_g \right)  + \mathcal{H}^{-2}  \partial_{\chi} \left( v_{||} \partial_{\chi} v_{||} \right) \,,
\\
\label{Delta2R}
\Delta^{(2)}_R &=-&\left( \frac{{\mathcal{H}'}}{\mathcal{H}^2} + \frac{2 - 5s}{\chi \mathcal{H}}+ 5s - b_e\right) v_{||}^{(2)}
+ \left( 1 + 3 \frac{{\mathcal{H}'}}{\mathcal{H}^2} + \frac{4 - 5s}{\chi \mathcal{H}} + 5s - 2 b_e \right) \mathcal{H}^{-1} v_{||} \partial_{\chi} v_{||}
\nonumber \\
&&
 - \left( \frac{{\mathcal{H}'}}{\mathcal{H}^2} + \frac{2 - 5s}{\chi \mathcal{H} } + 5s - b_e \right) v_{||} \delta_g  
+ \mathcal{H}^{-1} v_{||}  \delta_g' - 2 \mathcal{H}^{-1} v^a_{\perp} \partial_{a{\perp}} v_{||}
\nonumber \\
&&\label{kdsjncskjdvbskjdvbskj}
 -  \mathcal{H}^{-2} \psi \partial_{\chi}^2 v_{||} + \mathcal{H}^{-1} \psi \partial_{\chi} \delta_g + \mathcal{H}^{-2} v_{||} \partial_{\chi}^2 \psi \, ,
\eea
where $\psi$ 
\cc{is the gravitational potential, and $(2)$ superscript denotes the second-order part. We have followed the perturbative expansion convention $\Delta=\Delta^{(1)}+\Delta^{(2)}$ consistent with the expressions in \cite{Dio_2019}}\footnote{Our $\boldsymbol{n}$ is minus theirs.}. 
\cc{A  galaxy bias expansion can be introduced for the galaxy number count in real space,} 
\bea
\delta_g^{(2)} = b_1 \delta^{(2)} + b_2 \delta^2.
\eea
\cc{(We ignore the tidal bias for simplicity.)}
Equations (\ref{Delta2N}) and (\ref{Delta2R}) represent the second-order contributions to the observed galaxy number counts, split into Newtonian $\Delta^{(2)}_N$ and relativistic $\Delta^{(2)}_R$ components. We have only given the terms that scale as $\mathcal{H}/k$. However, the full expression has been calculated before \cite{Bertacca_2014} \cc{and is considerably more complicated}.  The Newtonian part contains nonlinear couplings between velocity and galaxy over-density, and scales as $(\mathcal{H}/k)^0$. In contrast, the relativistic components encapsulate the effects that scale as $\mathcal{H}/k$ and are particularly relevant on larger scales. The first set of terms involves coefficients dependent on the conformal Hubble parameter, evolution and magnification bias. Altogether, these terms give rise to the odd multipoles in higher-order statistics, such as the dipole in the bispectrum \cite{Clarkson_2019,PhysRevD.102.023530}. 

\subsection{General form of the relativistic redshift space distortions}

For the first-order Kaiser effect it is common to think of this as a point-wise distortion whereby the RSD generates a quadrupole (and monopole), and a circular pattern of galaxies gets distorted into an elliptical shape the via a $\mu^2$  correction, where $\mu= \bm n\cdot \hat{\bm v}$.\footnote{\cc{We remove the irrelevant azimuthal angle from our discussion throughout, so in reality we have a sphere being distorted into an ellipsoid etc. }}  
This becomes considerably more complicated once we go to higher order and include relativistic terms. In general we have,
\be
\Delta(z,\bm n) = \sum_{\ell} \Delta_{\ell}(z)\mathcal{L}_\ell(\mu)\,,
\ee
where $\mathcal{L}_\ell(\mu)$ are the  Legendre polynomials. In addition,
\be
\Delta_{\ell}(z) = \sum_n \Delta^{(n)}_{\ell}(z)\,,
\ee
where $n$ represents the perturbative order. 
Then, the first order Newtonian term has non-zero $\Delta^{(1)}_{0}, \Delta^{(1)}_{2}$. The second-order Newtonian terms generate non-zero $\Delta^{(2)}_{0}, \Delta^{(2)}_{2}$ and $\Delta^{(2)}_{4}$.  In the 2-point correlation function this therefore generates even multipoles up to order 8. 
At each perturbative order $n$ higher than this we generate corrections which have higher multipoles, with $\ell_\text{max}(n) = \ell_\text{max}(n-1) + 2$, and hence are all even parity with respect to the line of sight. (That is, each perturbative order gets an extra $\mu^2$ dependence.) All of these multipoles arise from factors of $\bm n \cdot \bm v$ and line of sight derivatives $\bm n \cdot \bm\nabla$. (These multipoles are defined with respect to radial in-fall or outflows of matter about a given observation point $(z,\bm n)$~-- a generic peculiar velocity field will have a more complicated multipole structure.)

For the most part the relativistic picture is similar but with each $\ell$ odd rather than even for the dominant $\mathcal{H}/k$ terms we consider here. This is responsible for breaking the symmetry along the line of sight -- the relativistic terms  swap sign under $\bm n\to-\bm n$. So we have at linear order non-zero $\Delta^{(1)}_{1}$, and at second-order we introduce a non-zero octupole $\Delta^{(2)}_{3}$, at third-order we have $\Delta^{(3)}_{5}$ etc. Again all of these multipoles arise from factors of $\bm n \cdot \bm v$,  $\bm n \cdot \bm\nabla$ and $\bm v_\perp\cdot\bm\nabla$. 

Although we will not go into the details here the relativistic picture is more complicated than this. There are many additional local  contributions at higher powers of $\mathcal{H}/k$, up to $(\mathcal{H}/k)^4$ at second-order (e.g., from $\psi^2$ terms etc.). At second order, even powers of $\mathcal{H}/k$ contribute to the even multipoles, so the Newtonian terms receive a correction, including the monopole (which is important for $f_\text{NL}$ measurements for example). Similarly odd powers contribute to the odd multipoles. Furthermore, there are integrated contributions which give rise to a multipole structure which is non-local. 

Although the multipole structure of this analysis is mathematically natural in terms of Legendre polynomials (as it is the non-trivial part of a spherical harmonic decomposition), for visualising the relativistic contributions this is not the only relevant basis. For example, in~\eqref{kdsjncskjdvbskjdvbskj} we can read off
powers of $\mu$ directly from counting powers of $\bm n \cdot \bm v$ and  $\bm n \cdot \bm\nabla$, so this is also a natural~-- but not orthogonal~-- way to view the higher-order contributions. Alternatively, a Fourier series basis on the circle using Chebyshev polynomials of the first kind $T_\ell(\cos\theta)=\cos(\ell\theta)$ (where $\mu=\cos\theta$) is also a useful orthogonal basis for expanding and visualising the relativistic contributions.\footnote{The first 4 Chebyshev polynomials are
\[
\begin{aligned}
T_{0}(\mu) &= 1
          \;=\; \mathcal{L}_{0}(\mu), ~~~~
T_{1}(\mu) = \mu
          \;=\; \mathcal{L}_{1}(\mu), ~~~~
T_{2}(\mu) = 2\mu^{2}-1
          \;=\; \frac{1}{3}\!\bigl(4\,\mathcal{L}_{2}(\mu)-\mathcal{L}_{0}(\mu)\bigr), \\
T_{3}(\mu) &= 4\mu^{3}-3\mu
          \;=\; \frac{1}{5}\!\bigl(8\,\mathcal{L}_{3}(\mu)-3\,\mathcal{L}_{1}(\mu)\bigr).
\end{aligned}
\]} 
\begin{figure}
    \centering
    \includegraphics[width=0.95\linewidth]{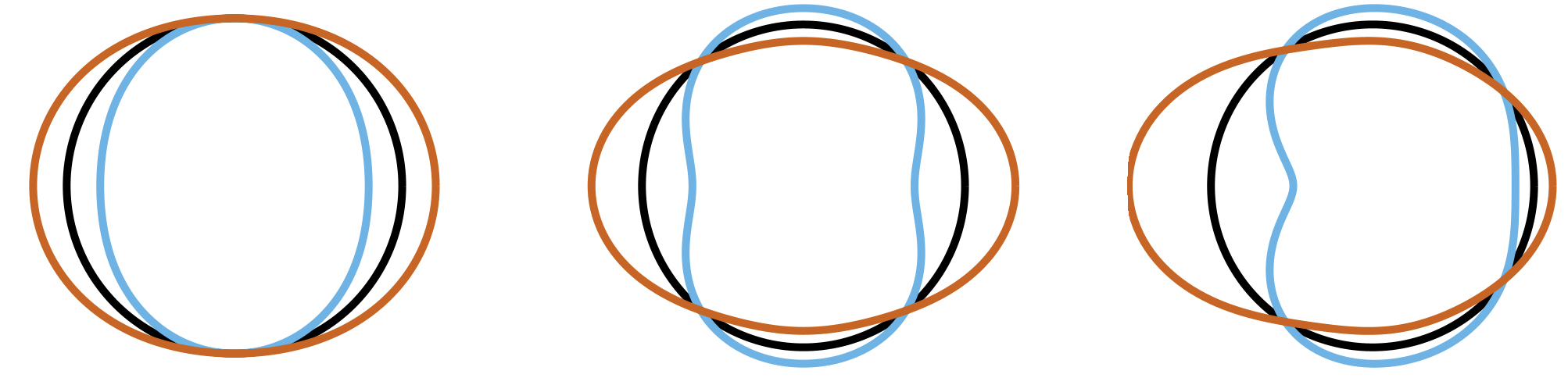}
    \caption{A sphere of particles (black in real space) either in-falling (blue) or out-flowing (brown) from the centre, seen by an observer at the left in redshift space.
    The left shows just the Kaiser distortion for reference~-- i.e., this is a polar curve with  $r(\theta)=1+\epsilon\cos^2\theta$ which is approximately an ellipse for $|\epsilon|\ll1$. 
    The other two show the sum of contributions for a Legendre series up to $\ell=4$. The middle plot consists of even contributions, mimicking Newtonian contributions, and right also contains odd multipoles, thereby adding in relativistic corrections (vastly exagerated). The curves have even $\ell$ positive (blue) or negative (brown), which show the different distortions that can apply depending on the sign of $\bm v$. The odd $\ell$ on the right have the symmetry broken into an egg or a bean shape depending on the signs of each contribution. The relative size of each multipole has been set arbitrarily for clarity with the odd multipoles a smaller contribution than the even.}
    \label{fig:enter-label}
\end{figure}

\begin{figure}
     \centering
     \includegraphics[width=1\linewidth]{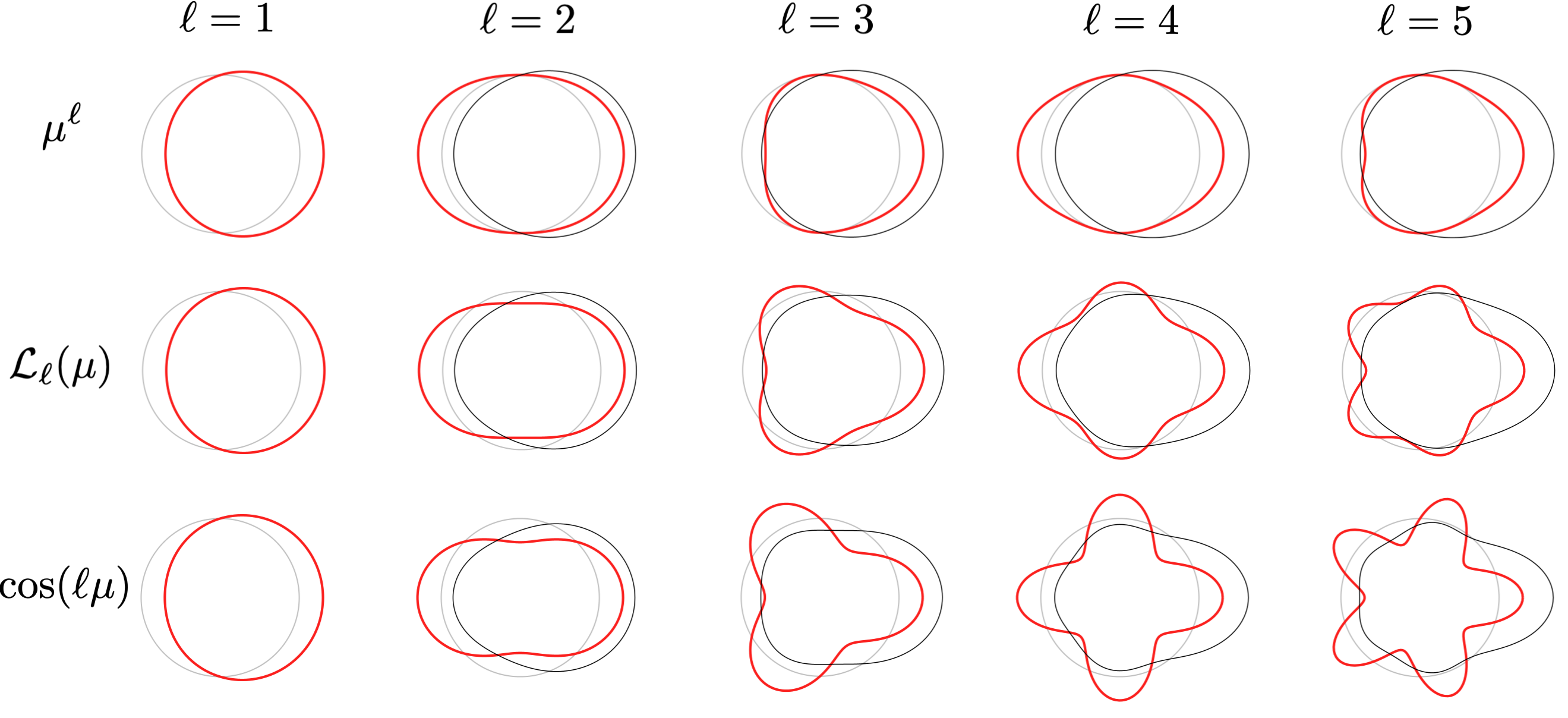}
     \caption{Illustration of the distortion of a small ring of particles from a circle (grey). This is for the three simplest ways of decomposing the distortions at each `multipole' order $\ell$. In red is a 30\% distortion (e.g., the plot in polar coordinates is $r(\theta)=1+0.3 \mu^\ell$ in the first row), and in black is the accumulated distortion with each multipole suppressed by a factor of $\ell$.  }
     \label{fig:circle-distortions}
 \end{figure} 
 
Let us assume we have a small ring of particles which are radially flowing inwards or outwards about an observation point. Consider the effect of a idealised small distortion to the this circle using $r(\theta)=1+\varepsilon_\ell\mathcal{L}_\ell(\mu)$ in the Legendre polynomial case, and similarly for the other expansion bases. This would apply to a small circle of galaxies which have an isotropic velocity field, viewed from the left. In Fig.~\ref{fig:enter-label} we show the sum of the contributions from odd and even multipoles depending on the relative signs of each contribution. In this example we consider $\varepsilon_\ell$ the same sign or alternating, giving rise to egg or bean shapes in the relativistic case. For reference we also give the standard Kaiser distorion in the left panel of this figure. 

Let us now consider each type of contribution separately, for each type of basis function. 
For $\ell=1$ we have a dipolar centroid shift regardless of  basis function as they are the same. (Note that these perturbations are not area preserving at $O(\varepsilon^2)$, so it also increases the size.) For $\ell=2$ we have elliptical-like distortions. For the $\mu^2$ case it's a pure ellipse corresponding to the the Kaiser term with the orientation depending on the sign of $\varepsilon_2$. 
When decomposed into orthogonal functions part of this is transferred to a monopole which results in a squashed ellipse for the quadrupole~-- compare $\mathcal{L}_2$ and $T_2$. The result of the dipole and quadrupole results in an egg-shaped distortion (we have removed the monopole perturbation in the sum which means they are not quite the same). Higher multipoles add extra features to the resulting distribution and are symmetric along the line of sight if $\ell$ is even, and antisymmetric if $\ell$ is odd.

\section{Idealised structures in redshift space} \label{sec3}

We will now look at some idealised structures in redshift space. We will assume a spherically symmetric over- or under-density, and consider how they appear in a plane slicing through the centre of symmetry. We define the line of sight direction to correspond to an observer looking from the left in all the figures, i.e., along the $x$-axis, $\boldsymbol{n} = \hat{\bm e}_x$, where 
\be
\hat{\bm e}_x=\cos \theta \hat{\bm e}_r - \sin  \theta \hat{\bm e}_{\theta}.
\ee
Here, $r$ and $\theta$ are polar coordinates from the centre of the object located a distance $\chi$ from the observer (who is located at $x=-\chi, y=0$). 

Given the gravitational potential of the over- or under-density, 
we have the peculiar velocity in terms of the gravitational potential,
\begin{equation} \label{pecvel}
    \boldsymbol{v}(r) = -\frac{2f}{3 \Omega_m Ha^2} \nabla \psi({r}). 
\end{equation}
Then, the radial component of the velocity, $\boldsymbol{v} \cdot \boldsymbol{n}$, is given by 
\bea
v_{||} = \boldsymbol{v}\cdot \boldsymbol{n} = - \frac{2f}{3 \Omega_m Ha^2}\cos\theta \partial_r \psi. 
\eea
Derivatives along the line of sight of the observer, which appear in $\Delta$ are then 
\be
\partial_\chi=\bm n\cdot\bm \nabla = \cos\theta\partial_r - \frac{1}{r}\sin\theta\partial_\theta\,.
\ee
Then we find, for example the Kaiser term becomes
\begin{equation}
   \partial_{\chi} v_{||} =  \boldsymbol{n}\cdot \bm\nabla(\boldsymbol{v} \cdot \boldsymbol{n}) =-\frac{2f}{3\Omega_m Ha} \Big(\cos^2\theta \partial^2_r \psi + \frac{1}{r} \sin^2\theta \partial_r \psi \Big).
\end{equation}
Other more complicated terms in $\Delta$ follow in a similar way. We choose the centre of the object at $z=1$ and a universe with $H_0=67$km\,s$^{-1}$Mpc$^{-1}$ and $\Omega_m=0.3$.



\subsection{The Isothermal Sphere}


As a simple model of an over-density we will model the density profile as a (non-singular) pseudo-isothermal sphere, 
\bea
\rho(r) - \bar\rho = \rho_0 \left[ 1 + \left(\frac{r}{r_0} \right)^2 \right]^{-1},
\eea
where $r$ is the radial distance from the centre. Here, $\rho_0$ is the finite central density and $r_0$ is the characteristic radius. These parameters depend on the properties of individual halos. The gravitational potential is calculated from the Poisson equation,
\begin{align}
\psi &= \frac{1}{r} \int_0^r \mathrm{d}r'\int_0^{r'}\mathrm{d}r'' r'' 4\pi G \rho_0 \left[1 + \left( \frac{r''}{r_0}^2 \right)\right]^{-1},  \\ \nonumber
&= \frac{3 \Omega_m H^2 r_0^2\rho_0}{2r_b\bar{\rho}}\left[r_b\mathrm{ln}(r_b^2 + 1) + 2 \mathrm{arctan}(r_b) - 2r_b \right],
\end{align}
where $r_b = r/r_0$. (We set the potential to be zero at the centre as the potential is infinite as $r\to\infty$, an unrealistic but irrelevant aspect of this model.) From (\ref{pecvel}), the velocity profile is given by,
\begin{align}
    v(r) = \frac{2fH\rho_0}{a^2r_b^2\bar{\rho}} \big\{r_0 \left[r_b - \mathrm{arctan}(r_0)\right] \big\}.
\end{align}
\begin{figure} [h]
    \centering
    \includegraphics[width=1\linewidth]{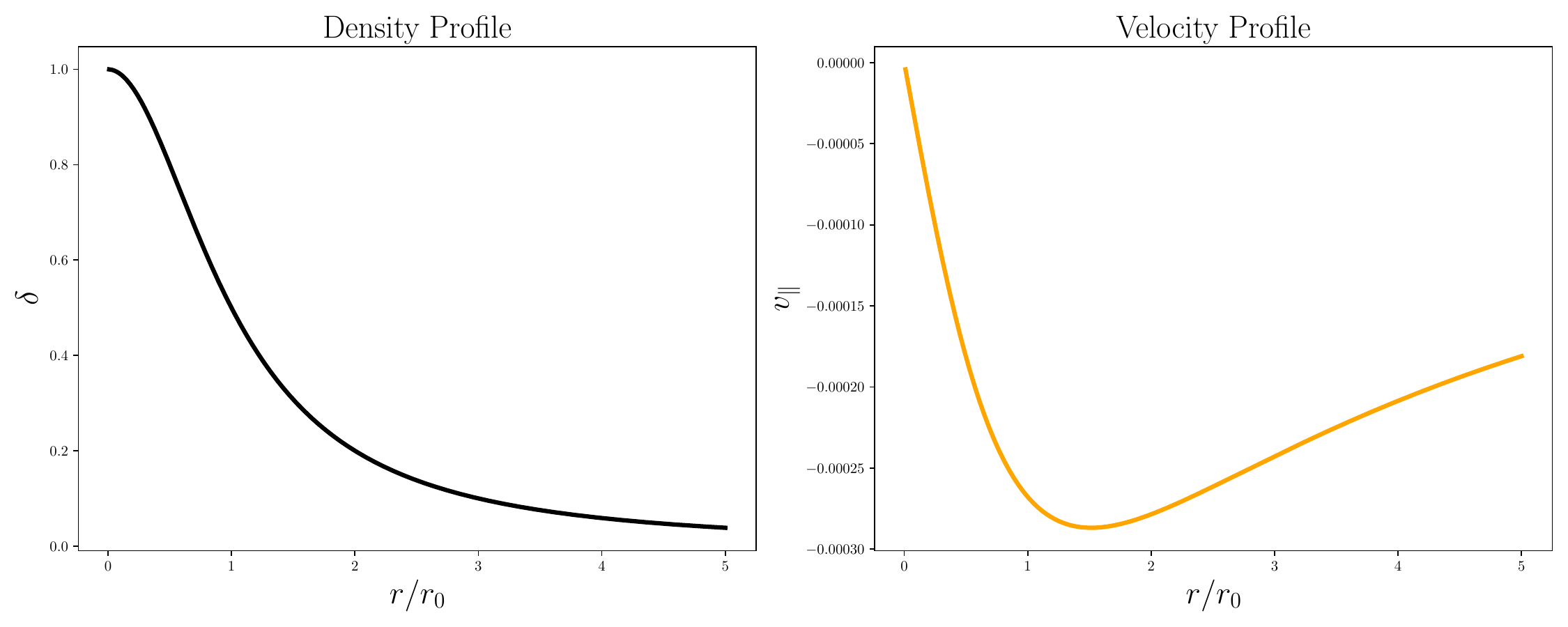}
    \caption{Radial profiles for the isothermal sphere. The left plot shows the density profile, which falls off with radius. The right plot displays the peculiar velocity profile, which exhibits a minimum and approaches $0$ for large radii. \cc{We set $\delta=1$ at the origin.} 
   }
\label{fig:1Disosphereprof}
\end{figure}

\begin{figure}[h]
    \centering
    \includegraphics[width=1\linewidth]{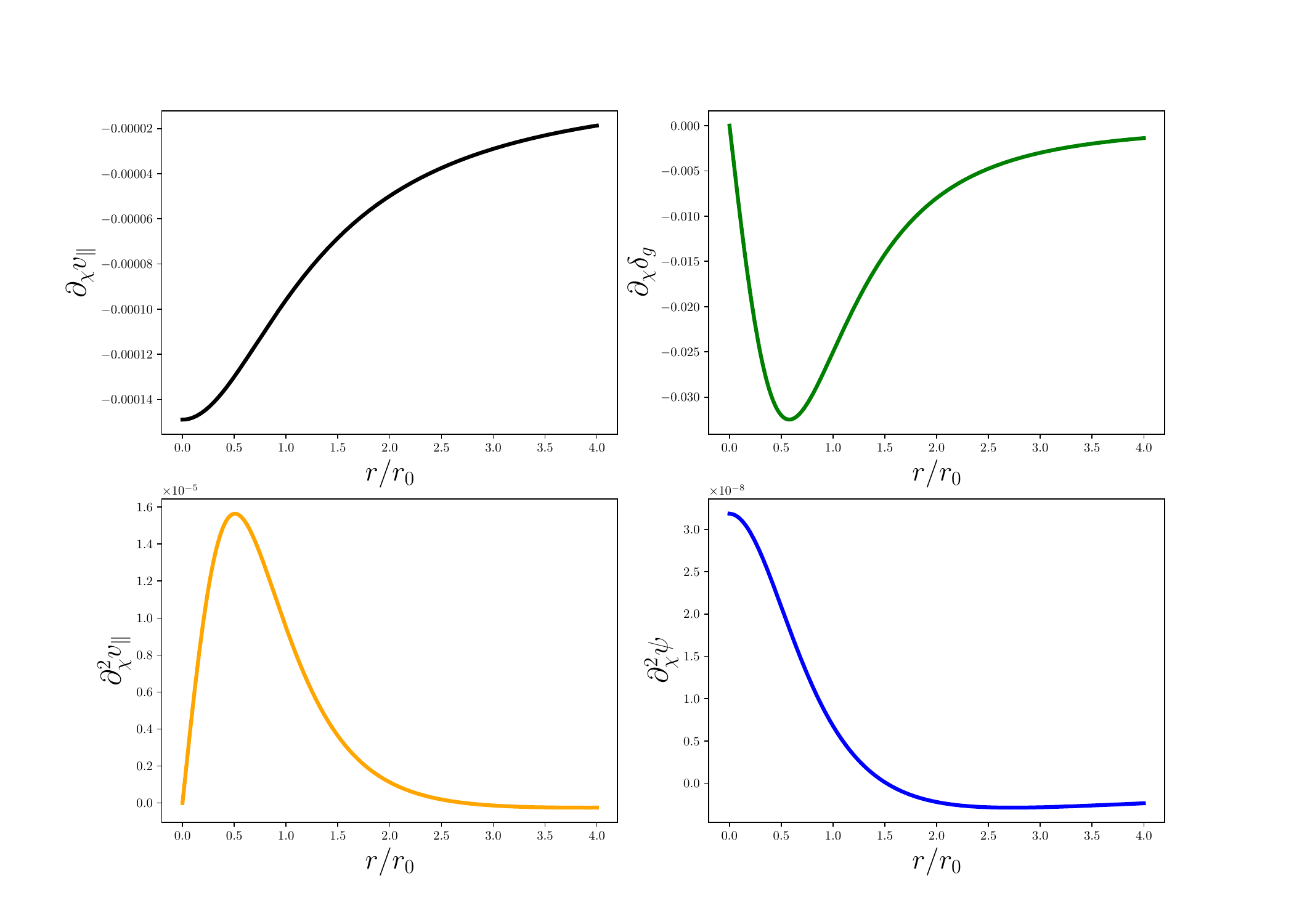}
    \caption{The top-left plot shows the radial derivative of the peculiar velocity, $\partial_{\chi}v_{\parallel}$, which gradually increases in amplitude with the radius. The top-right plot displays the radial derivative of the galaxy density contrast, $\partial_{\chi}\delta_g$, with a minimum at low radius. The bottom-left plot shows the second derivative of the peculiar velocity, $\partial^2_{\chi}v_{\parallel}$, which peaks at low radii and falls off after. The bottom-right plot presents the second derivative of the gravitational potential, $\partial_{\chi}^2\psi$, which decreases with radius. Note: The $x$-axis is the same as in Fig.~\ref{fig:1Disosphereprof}.}
    \label{fig:1Disosphereprof12}
\end{figure}

Fig.~\ref{fig:1Disosphereprof} shows the radial profiles for the isothermal sphere. The density profile (on the left) decreases smoothly with radius, while the velocity profile (on the right) shows in-fall with a minimum before tapering off at large radii. 
\cc{For ease of reference, } Fig.~\ref{fig:1Disosphereprof12} shows the line of sight derivatives used in redshift-space distortion terms for the isothermal profile. 

Clearly, using an isothermal sphere doesn't model a cluster near the origin because the radial velocity approaches zero. Of course in this regime a realistic cluster would be virialised and Fingers of God would appear near the centre. A more realistic profile such as Navarro-Frenk-White (NFW) has a divergent density at the origin, which is also unrealistic, but also makes the subtle effects we are looking at hard to visually pick out. We are interested in the region $r/r_0\gtrsim1$, which is where the relativistic terms are relevant, and is where the isothermal sphere is a reasonable approximation. Although the velocities are lower in this region rather than in the heart of a cluster, the relativistic terms are relevant relative to the Hubble scale, so this is the regime of interest.

\begin{figure} [h]
    \centering
    \includegraphics[width=0.99\linewidth]{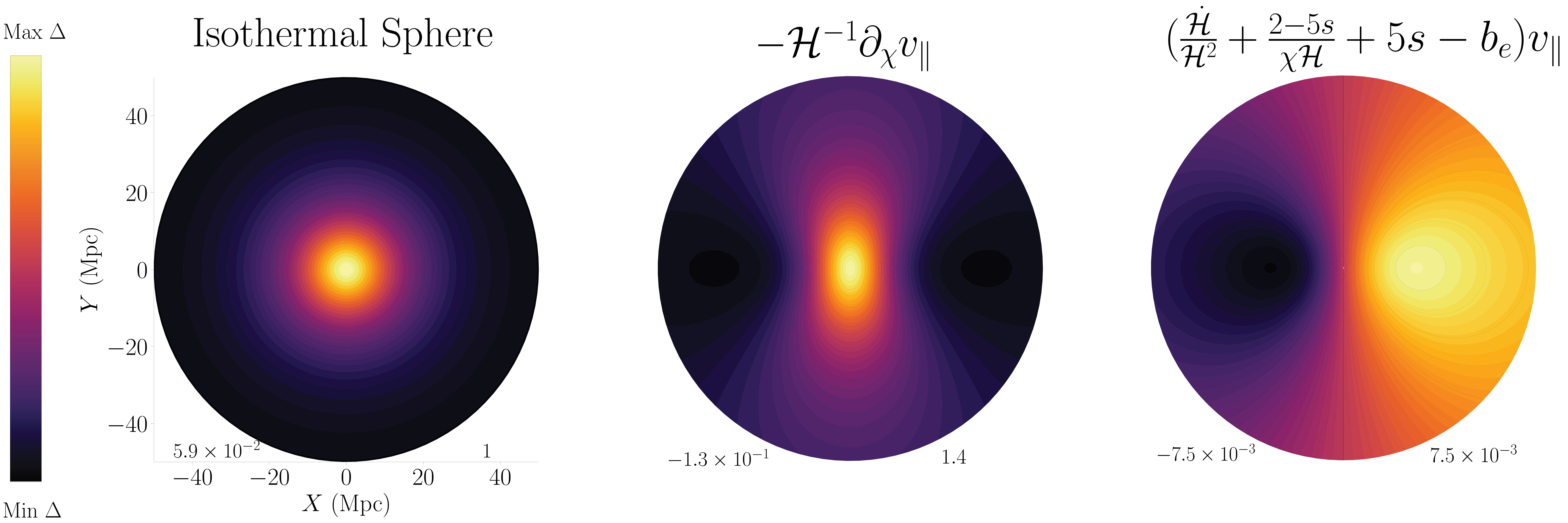}
    \caption{First order contributions to galaxy number counts in an isothermal sphere profile. The left plot shows the spherically symmetric density contrast of the isothermal sphere, with a central over-density that gradually falls off with radius. The middle plot displays the Newtonian redshift-space distortion term, which introduces a symmetric pattern along the line of sight. The right plot is the relativistic Doppler contribution, which has a dipolar asymmetry along the line of sight.}
    \label{fig:First_order_terms_Iso_sphere}
\end{figure}

In Fig.~\ref{fig:First_order_terms_Iso_sphere}, we show the spherically symmetric isothermal over-density as a 2-d density plot. The left plot of Fig.~\ref{fig:First_order_terms_Iso_sphere} displays the real space density profile. This profile effectively serves as the background matter distribution for visualising RSD. The middle plot shows the first order Kaiser term defined in (\ref{Delta1N}), which is $-\mathcal{H}^{-1}\partial_{\chi}v_{\parallel}$. This term is referred to as the standard redshift-space distortions and leads to a symmetric compression of the density field along the line of sight, transforming spherical structures into ellipses. The right plot illustrates the first order Doppler term defined in (\ref{Delta1R}), which stems from relativistic corrections to galaxy number counts. Unlike the Kaiser term, this contribution is directly proportional to the line of sight peculiar velocity. The amplitude of this term encodes information about time evolution through parameters such as the magnification bias, $s$, and evolution bias, $b_e$.  This term is suppressed by a factor of $\mathcal{H}/k$ relative to the Kaiser term, making it subdominant and exhibiting a characteristic dipolar pattern.

\begin{landscape}
\begin{figure}[p]
    \centering
    \includegraphics[width=0.99\linewidth]{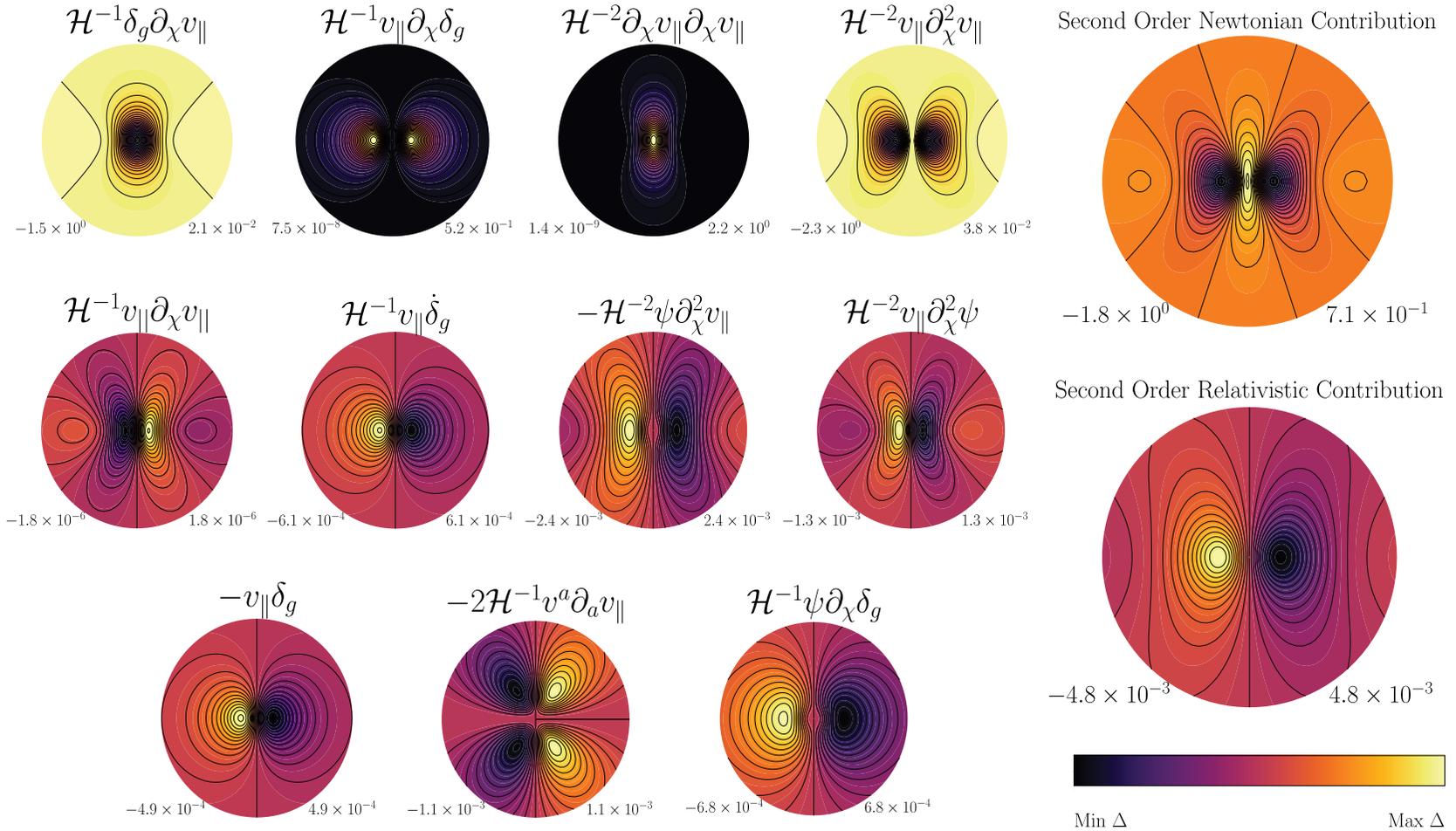}
    \caption{Second order contributions to galaxy number counts in an isothermal sphere profile. The top row shows the individual Newtonian second-order terms. The top right panel displays the total second-order Newtonian contribution, which has symmetric contributions along the line of sight. {The second and third row are representative of the individual relativistic terms.}  The bottom right panel shows the total second-order relativistic contribution, which is asymmetric along the line of sight, in the form of a dipolar structure. \cc{Note that the numbers refer to the minimum and maximum the values take in the figure, and as we have chosen $\delta_0=1$ these can be thought of in units of $\delta_0^2$.} }
    \label{fig:second_order_terms_Iso_sphere}
\end{figure}
\end{landscape}

To explore the second-order perturbations in relativistic galaxy number counts, we plot the contributions from both Newtonian and relativistic terms, visualised for the same isothermal overdensity. Fig.~\ref{fig:second_order_terms_Iso_sphere} decomposes the contributions into Newtonian, shown in the top row, \cc{and} the relativistic contributions are depicted in the middle and last rows. The combined total Newtonian and relativistic signals are shown in the far-right plots. 
The Newtonian terms, corresponding to (\ref{Delta2N}), consist of combinations of first order density and velocity fields and their derivatives. These produce distortions that are symmetric under line of sight reversal. Specifically, the terms $\delta_g\partial_{\chi}v_{\parallel}$ and $v_{\parallel}\partial_{\chi}\delta_g$ are proportional to $\cos^2 \theta$ while $\partial_{\chi}v_{\parallel}\partial_{\chi}v_{\parallel}$ and $v_{\parallel}\partial_{\chi}^2v_{\parallel}$ are proportional to $\cos^4\theta$. The term, $\delta_g\partial_{\chi}v_{\parallel}$, couples the galaxy over-density, $\delta_g$, and the line of sight gradient of the peculiar velocity, $\partial_{\chi}v_{\parallel}$. 
It exhibits an under-density at the centre and the contours are elongated vertically. The term, $v_{\parallel}\partial_{\chi}\delta_g$, shows a symmetric bi-lobed pattern, with over-dense lobes near the centre that gradually transition to under-densities as we move radially outward. 
The third plot, representing $\partial_{\chi}v_{\parallel}\partial_{\chi}v_{\parallel}$, corresponds to the square of the velocity gradient term. The term is always positive and peaks at the centre and drops off as we move away from the centre. 
Lastly, $v_{\parallel}\partial^2_{\chi}v_{\parallel}$ also has a bi-lobed shape which is again invariant under $\boldsymbol{n} \rightarrow -\boldsymbol{n}$. It is under-dense in the centre and becomes more dense as we move towards the edge. 

The remaining plots in Fig.~\ref{fig:second_order_terms_Iso_sphere} correspond to the terms in (\ref{Delta2R}). These terms involve combinations of line of sight peculiar velocities, gravitational potentials, and their derivatives. In contrast to the Newtonian terms, these relativistic contributions break symmetry under line of sight reversal.
The first term, $v_{\parallel}\partial_{\chi}v_{\parallel}$, shown in the second row, is the Doppler $\times$ Kaiser term. It displays a prominent dipolar pattern, with over densities on one side and under densities on the other. \pritha{These overdensities indicate regions where number of particles are being shifted into due to peculiar velocities, reducing particles from the opposite side.} 
Like the first order relativistic  term in (\ref{Delta1R}), this term also encodes redshift evolution information through the astrophysical biases. \pritha{Importantly, the sign of the amplitude of the term can change depending on the biases, which can flip the direction of the dipole in the plot.} 
The plots for $\psi\partial^2_{\chi}v_{\parallel}$ and $v_{\parallel}\partial^2_{\chi}\psi$ also have a similar pattern, with over and under densities on the left and right-hand side, respectively. All the above terms are proportional to $\mathrm{cos}^3 \theta$. The terms, $v_{\parallel}\dot{\delta}_g$, $v_{\parallel}\delta_g$ and $\psi \partial_{\chi}\delta_g$ are proportional to $\mathrm{cos} \theta$ and exhibit the same general behaviour. Lastly, $v^a\partial_av_{\parallel}$ is the transverse Doppler term, which has no radial contribution. 

To the far right of Fig.~\ref{fig:second_order_terms_Iso_sphere}, on the top, we have the total contribution from the second order Newtonian terms. The plot shows multiple vertical lobes with an over density in the centre coupled with two under densities on either side. \pritha{The terms proportional to $v_{\parallel}\partial_{\chi}\delta_g$ and $\partial_{\chi}v_{\parallel}\partial_{\chi}v_{\parallel}$ dominate the Newtonian contribution, as evident from the structure of the plots. The centre of the total contribution is overdense with underdense lobes, which aligns with the patterns in the two dominating terms. In contrast, the other two terms contribute less prominently, both in shape and amplitude.} 
In the bottom plot, we show the second-order terms that include effects from gravitational potentials and peculiar velocities. Among these, $\psi\partial^2_{\chi}v_{\parallel}$ is the dominant contribution and is responsible for the characteristic shape of the plot. 

\begin{figure} [h]
    \centering
    \includegraphics[width=0.99\linewidth]{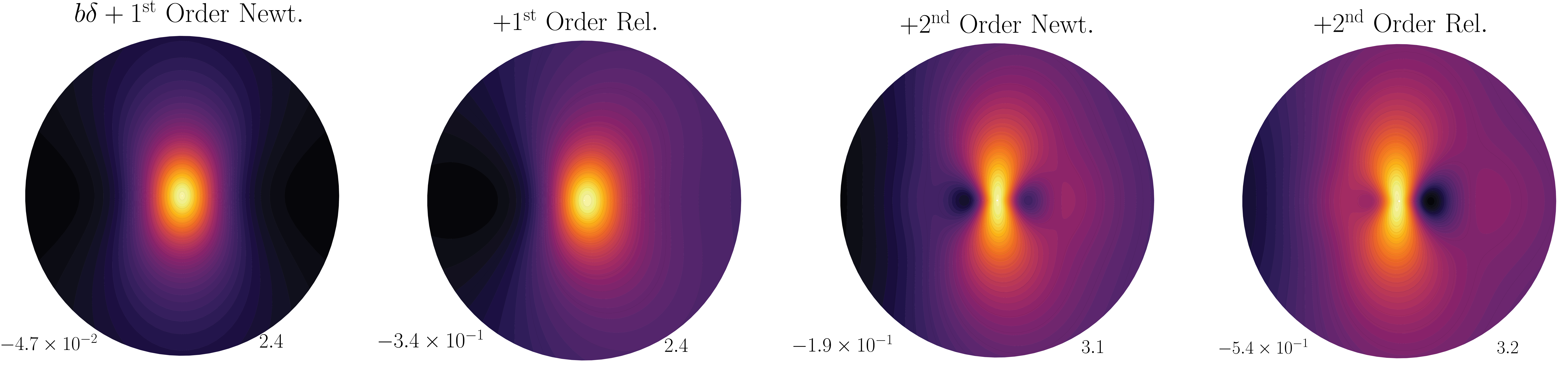}
    \caption{Cumulative number count contributions for an isothermal sphere profile. the leftmost plot adds the first-order Newtonian redshift-space distortion term. The second plot includes the Doppler term (magnified by $50$). The third plot shows all second-order Newtonian contributions, adding nonlinear structure and the contours become more tight in the centre. The final plot includes all relativistic corrections (the second-order relativistic terms are amplified by a factor of $200$ for clarity).}
\label{fig:total_cont_terms_Iso_sphere}
\end{figure}
Fig.~\ref{fig:total_cont_terms_Iso_sphere} illustrates the cumulative impact of first order and second order RSD and Doppler contributions. The first plot shows the effect of the first order Newtonian term on the density field. This agrees with literature where the initial spherical overdensity is squashed into an ellipse-like shape along the line of sight due to the peculiar velocities. The second plot introduces the Doppler effect at first order, leading to a clear asymmetry in the distribution, with one side appearing more over dense than the other. This effect also results in a centroid shift. The Doppler contribution, being suppressed by a factor of $\mathcal{H}/k$ relative to the Newtonian term, has been scaled by a factor of $50$ for visual clarity.

The third plot builds on this by adding in the second order Newtonian contribution. While the asymmetry from the first order Doppler effect remains visible, the inclusion of the non-linear Newtonian corrections introduced additional structure. In particular, there are quadrupolar patterns introduced due to terms being proportional to $\mathrm{cos}^2\theta$ and $\mathrm{cos}^4\theta$. The last plot shows the fully combined number counts, including all first and second order Newtonian and relativistic contributions, where the second order relativistic terms have been magnified by $200$ to enhance their visibility. Asymmetries are now more pronounced and are clearly visible through the contour lines. While the centre remains overdense, the radial profile becomes asymmetric. 

\subsection{Cosmic Voids}

The void density profile characterises the spherically averaged relative deviation of the mass density surrounding a void centre from the mean cosmic density. In \cite{Hamaus_2014}, the authors have proposed a simple formula to model this profile using tracer particles, by averaging their distribution in radial shells of thickness $2\delta r$, each located at a distance $r$ from the void centre. 
The density profile of a cosmic void is given by
\begin{equation}\label{densvoid}
    \delta(r) = \frac{\rho(r)}{\bar{\rho}} - 1 = \delta_c \frac{1 - (r/r_s)^{\alpha}}{1 + (r/r_v)^{\beta}},
\end{equation}
where $\delta_c$ is the central density contrast, $r_s$ represents a scale radius where $\rho_v = \bar{\rho}$. Here, $\alpha$ and $\beta$ tell us the slope of the inner and outer slope of the void's compensation wall respectively. Finally, $r_v$ is the effective radius of the void. In linear theory, the density contrast is related to the peculiar velocity, from which we define the radial velocity,
\begin{equation}
    v(r) = - \frac{1}{3} \Omega^{\gamma}_m Hr \tilde\delta(r),
\end{equation}
where $\Omega_m$ is the matter content in the Universe, $\Gamma$ is the growth index of matter perturbations, $H$ is the Hubble constant and $\tilde\delta(r)$ is the integrated density contrast defined as 
\begin{equation}
    \tilde\delta(r) = \frac{3}{r^3} \int^r_0 \Big(\frac{\rho_v(q)}{\bar{\rho}} - 1 \Big) q^2 \mathrm{d}q.
\end{equation}
With (\ref{densvoid}), the density contrast, $\tilde\delta(r)$, can also be expressed analytically using hypergeometric functions, 
\begin{equation}
    \tilde\delta(r) = \delta_c\, {}_2F_1\Big[1,\frac{3}{\beta},\frac{3}{\beta} + 1, -(r/r_v)^{\beta}\Big] - \frac{3 \delta_c(r/r_s)^{\alpha}}{\alpha + 3} {}_2F_1 \Big[1,\frac{\alpha + 3}{\beta},\frac{\alpha + 3}{\beta} + 1, -(r/r_v)^{\beta}\Big],
\end{equation}
where ${}_2F_1$ is the Gauss hypergeometric function. 

\begin{figure} [h]
    \centering
    \includegraphics[width=1\linewidth]{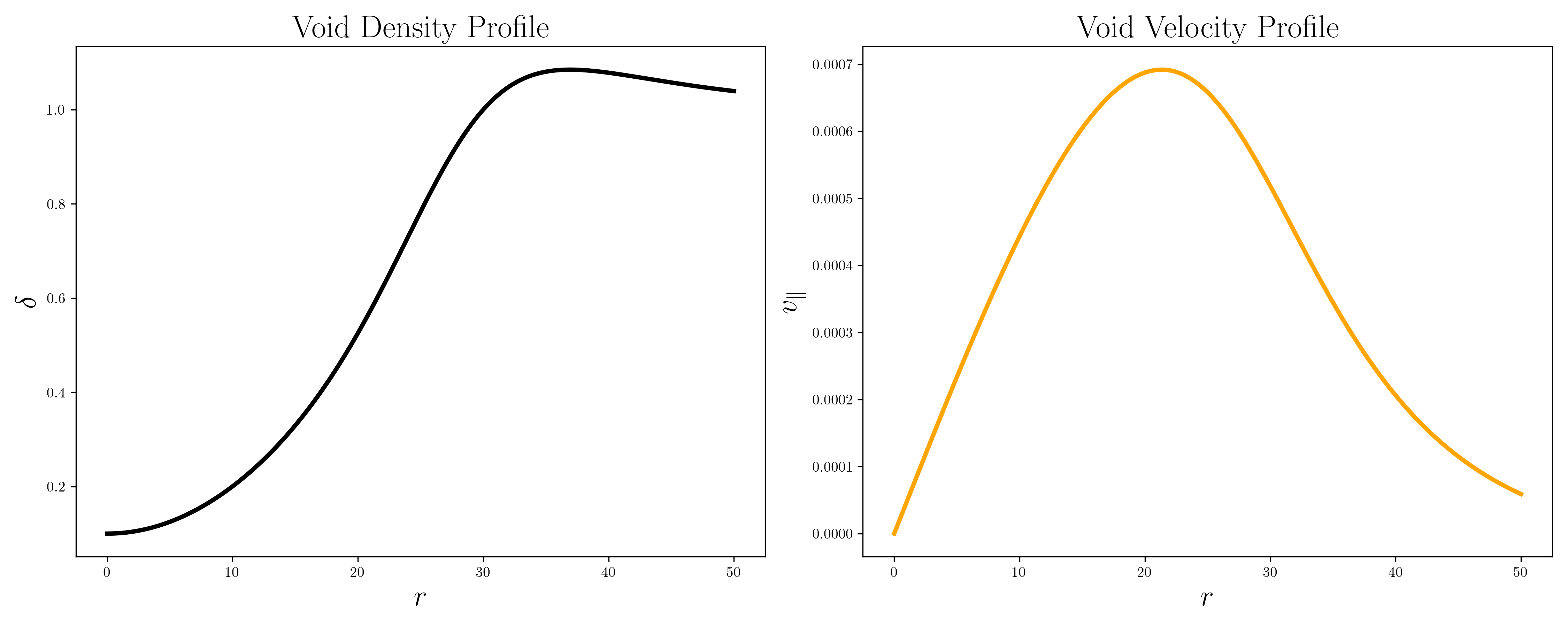}
    \caption{Radial profiles for the void model. The left plot shows the density contrast, which increases with radius, transitioning from an underdense centre to and overdense shell. This is in contrast to the isothermal sphere. The right plot shows the peculiar velocity profile, which rises to a peak and then falls off, indicating outward flows. In the isothermal case, the velocity profile exhibits infall.}
    \label{fig:1Dvoidprof}
\end{figure}

Fig.~\ref{fig:1Dvoidprof} illustrates the one-dimensional radial profiles of a void model. The left plot shows the radial density contrast, which increases with radius, and the corresponding peculiar velocity on the right plot. Unlike the isothermal sphere, the void exhibits outward flows. 

\begin{figure} [h]
    \centering
    \includegraphics[width=1\linewidth]{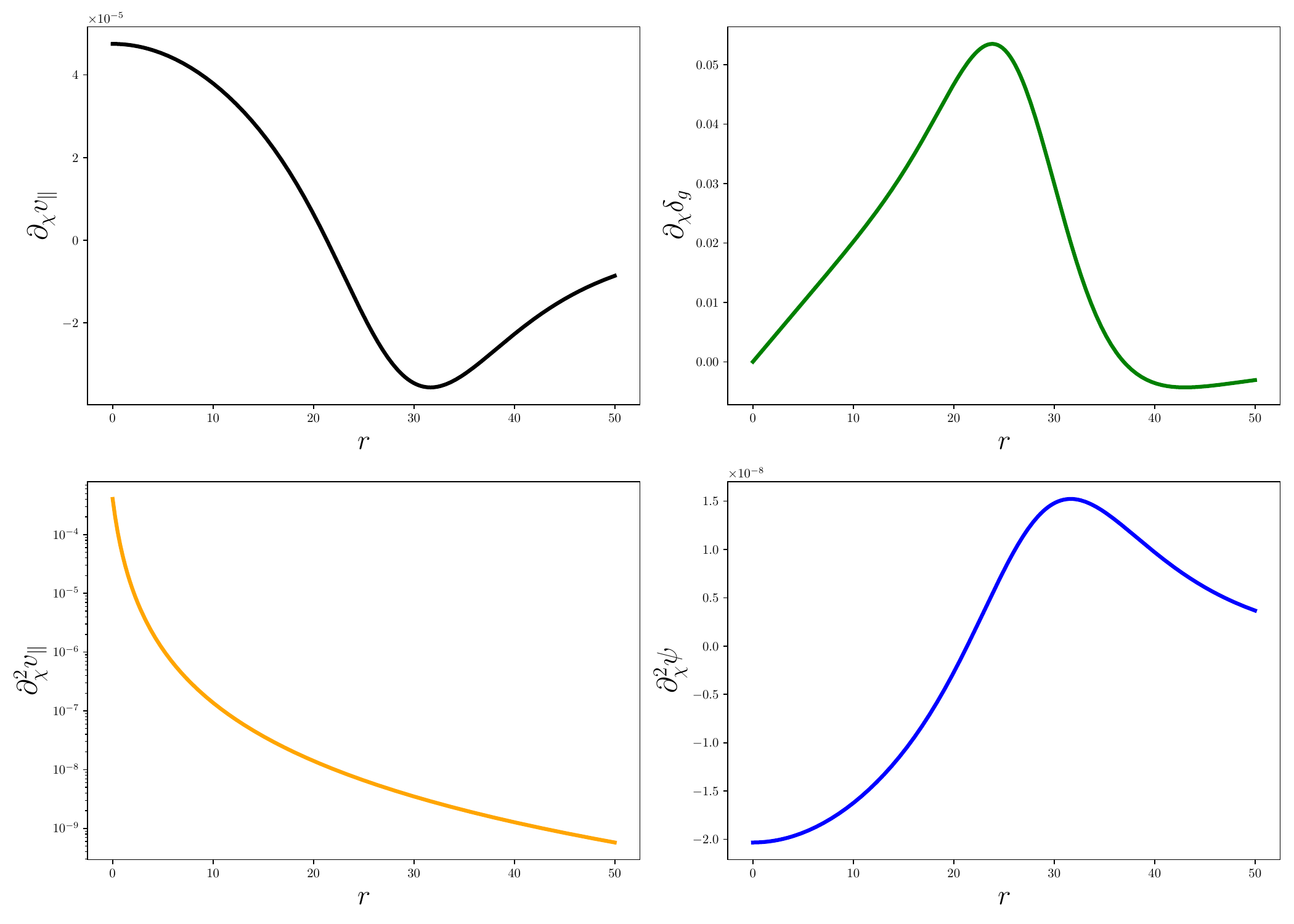}
    \caption{The top-left plot shows the radial derivative of the peculiar velocity, $\partial_{\chi}v_{\parallel}$, which exhibits a sign change across the profile. The top-right plot displays the radial derivative of the galaxy density contrast, $\partial_{\chi}\delta_g$, which peaks near the transition of the central underdensity and the surrounding overdensity. The bottom plots represent the second derivative of the peculiar velocity and the gravitational potential respectively. For the peculiar velocity, note the $y$ axis in an logarithmic scale to better visualise its decay with radius. Finally, $\partial^2_{\chi}\psi$ reaches a maximum and decreases towards the edges.}
    \label{fig:3Dvoidprof}
\end{figure}

Fig.~\ref{fig:3Dvoidprof} shows the radial derivatives of the key physical quantities. The derivative of the line-of-sight velocity, $\delta_{\chi}v_{\parallel}$, changes sign across the profile. The derivative of the density contrast, $\delta_{\chi} \delta_g$, exhibits a peak. The bottom plots are the second derivatives of the peculiar velocity and the potential.  

\begin{figure} [h]
    \centering
    \includegraphics[width=0.99\linewidth]{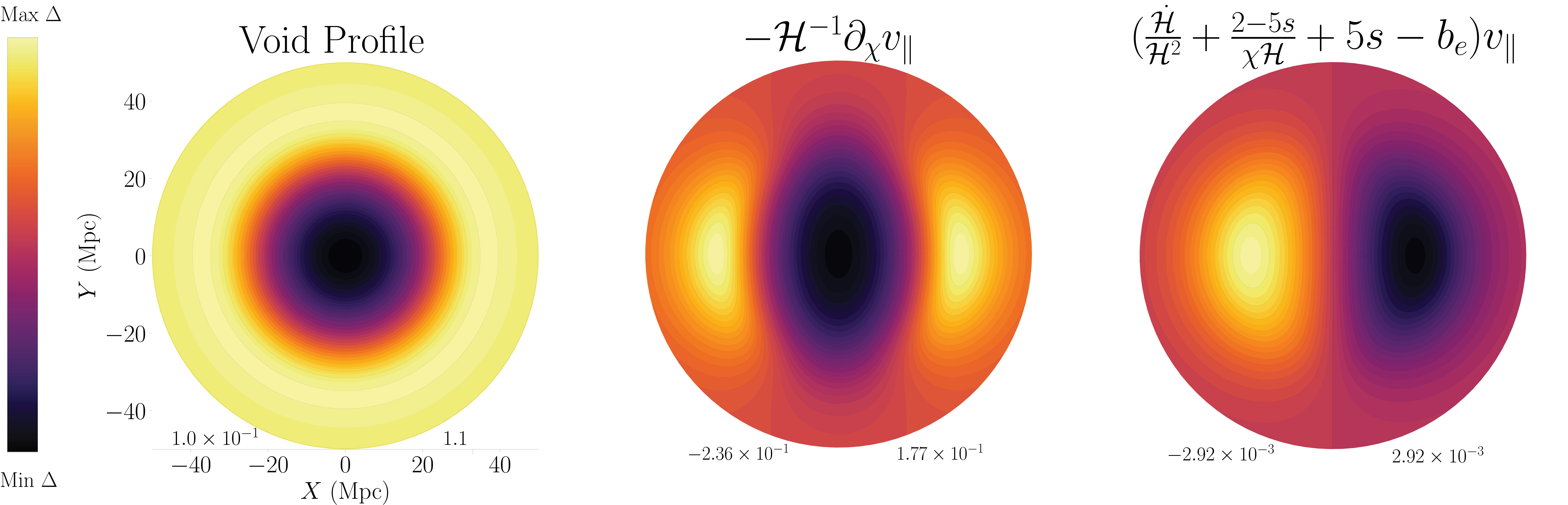}
    \caption{First order contributions to galaxy number counts in a void profile. The first plot shows the spherically symmetric density contrast. The middle plot displays the Newtonian redshift space distortion term, which is symmetric along the line of sight. The right plot shows the relativistic Doppler term, producing an asymmetric dipole with respect to the line of sight.}
    \label{fig:First_order_terms_Void}
\end{figure}

In Fig.~\ref{fig:First_order_terms_Void}, the first plot displays the density contrast, $\delta$, with a pronounced underdensity at the centre that increases radially outward. The middle plot shows the Kaiser term. In contrast to the isothermal sphere case shown in Fig.~\ref{fig:First_order_terms_Iso_sphere}, the void profile reveals a more prominent structure. There is an underdense centre followed by two distinct lobes of overdensity surrounding it. The final plot represents the Doppler term, which retains the characteristic dipolar pattern observed previously. 

\begin{landscape}
\begin{figure}[p]
    \centering
    \includegraphics[width=0.99\linewidth]{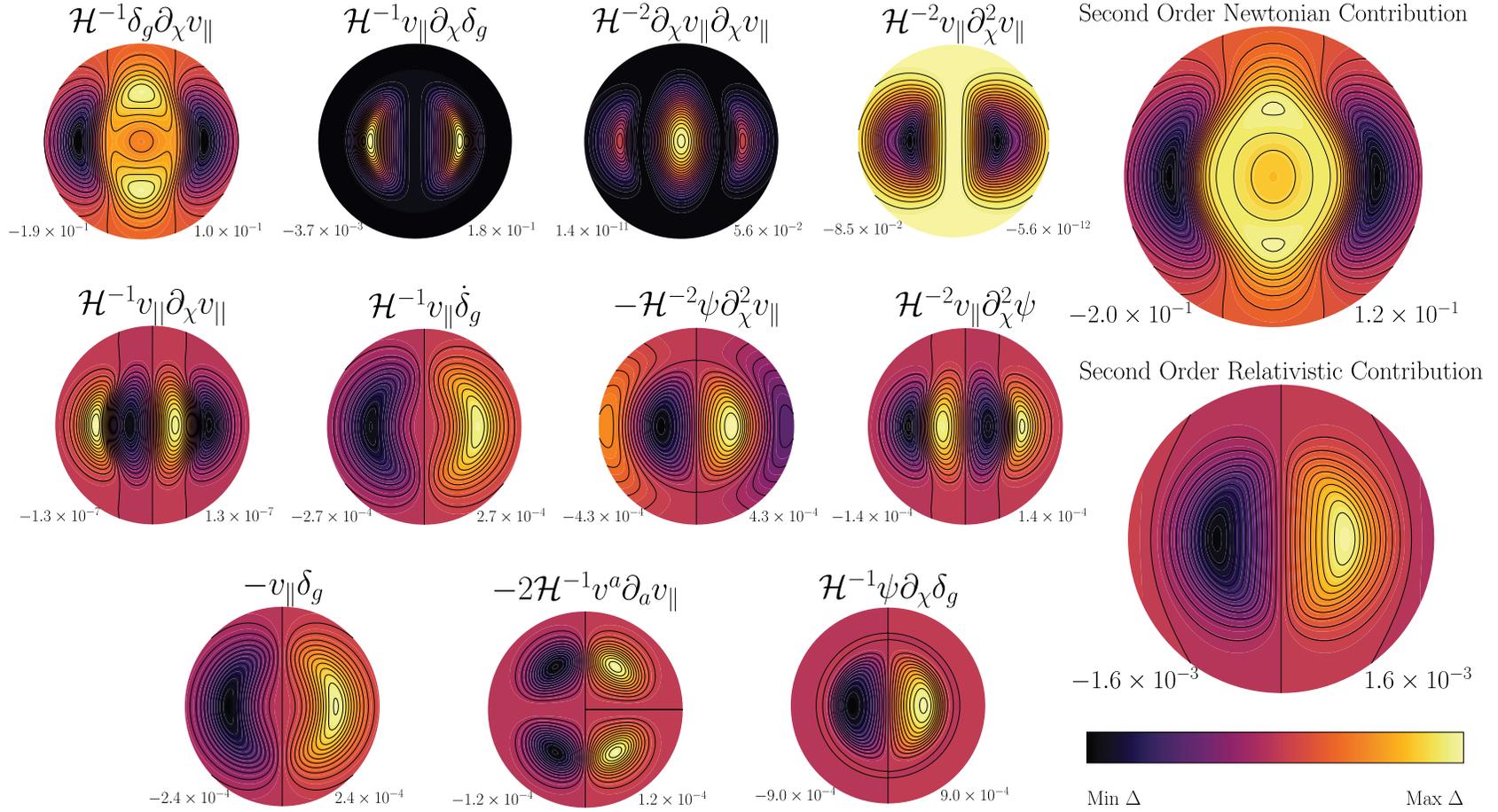}
    \caption{Second order contributions to galaxy number counts in a void profile. The top row shows the Newtonian terms, including couplings between density and velocity fields and higher-order velocity gradients. The final plot on the top right displays the total Newtonian contribution. The bottom two rows present the relativistic second-order terms, including Doppler term. The final plot on the bottom right shows the total relativistic contribution, with all the terms combined. Compared to the Newtonian terms, the relativistic contributions introduce strong asymmetries along the line of sight.}
    \label{fig:Second_order_terms_Void}
\end{figure}
\end{landscape}

The first row in Fig.~\ref{fig:Second_order_terms_Void} shows the terms in (\ref{Delta2N}) for a void profile, analogous to those shown previously for the isothermal sphere. The first plot, involving the product of the density contrast and the radial velocity gradient, exhibits a central overdense region accompanied by two underdense lobes, which are more prominent compared to the isothermal case. In the second plot, where the derivative of the density contrast is coupled with the radial velocity, we observe a pair of lobes characterised by central overdensities which become less dense as we move away from the centre. The third plot, $\partial_{\chi}v_{\parallel}\partial_{\chi}v_{\parallel}$, has three lobes. Unlike the case for the isothermal sphere, there is not only one central overdensity surrounded by underdense regions, but also two lobes on either side with a similar pattern. The last plot represents the term, $v_{\parallel}\partial_{\chi}^2v_{\parallel}$, which has two lobes with underdensities in the centre.

In Fig.~\ref{fig:Second_order_terms_Void}, the second and third rows correspond to the relativistic contributions. The first term, involving $v_{\parallel}\partial_{\chi}v_{\parallel}$, has four alternating lobes of positive and negative values. Compared to the isothermal sphere case, which only has two lobes, the void profile reveals more features. The terms, $v_{\parallel}\dot{\delta_g}$, $v_{\parallel}\delta_g$ and $\psi\partial_{\chi}^2v_{\parallel}$, both show asymmetric dipolar patterns, with the latter having an additional radial band with underdensity on the right and overdensity on the left. The fourth plot, $v_{\parallel}\partial_{\chi}^2\psi$, shows a pattern resembling that of $v_{\parallel}\partial_{\chi} v_{\parallel}$. The transverse Doppler term, $v^a\partial_av_{\parallel}$, has no contribution for $\theta = 0$ but there are four lobes in the four quarters with underdensities in the left quadrant and overdensities in the right. The term, $\psi\partial_{\chi}\delta_g$, has a sharper cut-off as we move away radially. This is due to $\psi$ going to $0$ at large distances from the void centre. To the right of the individual plots, there are the total contributions. The top right shows the total Newtonian contribution, with $\delta_g\partial_{\chi}v_{\parallel}$ and $v_{\parallel}\partial_{\chi}\delta_g$ dominating the shape and amplitude of the overall term. The bottom right shows the total relativistic contribution, with $v_{\parallel}\partial_{\chi}^2\psi$ and $v^a\partial_av_{\parallel}$ being the dominant terms.   

\begin{figure} [h]
    \centering
    \includegraphics[width=0.99\linewidth]{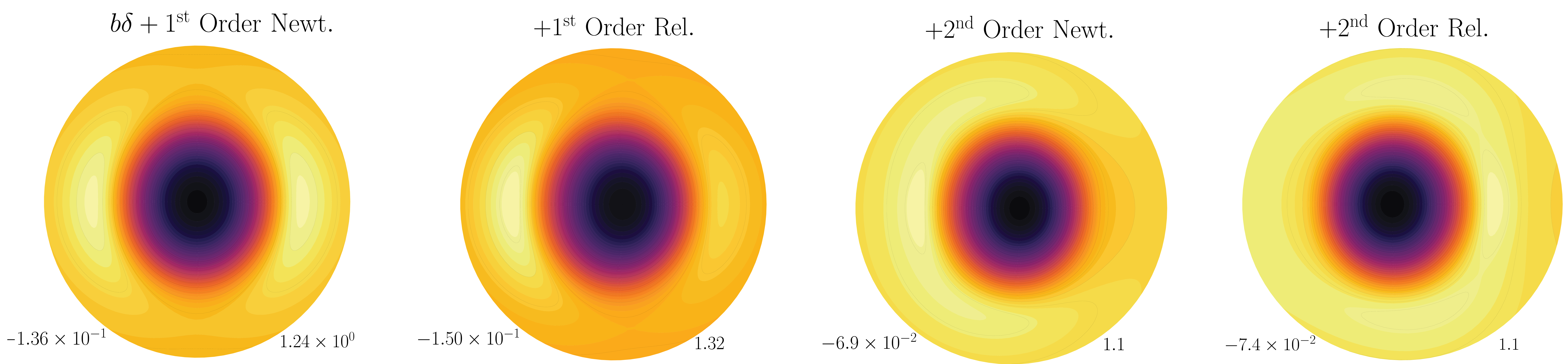}
    \caption{Cumulative contributions to galaxy number counts in a void profile. From left to right: the first plot shows the first-order Newtonian terms; the second adds the first order relativistic Doppler term (magnified by a factor of $50$). The third includes the full second-order Newtonian contributions; and the final plot shows the total corrections, combining all Newtonian and relativistic terms. The progressive asymmetry, shows the impact of relativistic effects, which distort the void structure along the line of sight.}
    \label{fig:total_conts_void}
\end{figure}

Fig.~\ref{fig:total_conts_void} presents the cumulative build-up of the contributions for a void profile. The first plot shows the first order Newtonian number count contributions, comprising the galaxy density contrast, $b\delta$, and the Kaiser RSD terms, $-\mathcal{H}^{-1}\partial_{\chi}v_{\parallel}$. The central region shows an underdensity surrounded by two symmetric overdense lobes. The second term has the first order Doppler term introduced, magnified by $50$. The term introduces an asymmetry, still having an underdense centre, with one side appearing more enhanced than the other. The third term incorporates the second-order Newtonian terms, which introduce couplings between density and velocity at second order. The central underdensity is stretched sideways forming an egg-like shape horizontally, with the asymmetry still present due to the first order Doppler term. The final plot includes all the Newtonian and relativistic corrections for the void profile, with the second order relativistic corrections magnified 200 times. Compared to the isothermal case, the asymmetry is different in the void profile, with tighter contour lines and more pronounced egg-like features, driven by the bump in the velocity profile that amplifies the relativistic corrections along the line of sight.  

\section{Conclusion}

In this paper, we have considered how simple over- and under-densities appear in redshift space when relativistic effects are included. In the Newtonian approximation, these distortions are all symmetric along the line of sight, giving rise to even multipoles in the density perturbation (and therefore in the 2pcf), with the maximum power of the multipole depending on the perturbative level considered. Relativistic effects, on the other hand, generate multipoles which are both symmetric as well as anti-symmetric along the line of sight, thereby fundamentally altering the picture of RSDs in general. 

In order to investigate this, we have considered the leading relativistic effects in terms of their amplitude in an $\mathcal{H}/k$ expansion, where they introduce only odd multipoles in redshift space. At linear perturbative order, this symmetry breaking means that a spherical infall or outflow of particles, which would be `elliptical' in redshift space in the Newtonian picture, now becomes egg-shaped or bean-shaped depending on the details of the flow. At higher order the distortions, though less pronounced, complicate this picture further. 
Showing this for idealised over- and under-densities illustrates in an intuitive way the manner in which each type of term distorts the pattern of galaxies an observer would see. 

One immediate application of analysing these effects in the observed density field would be when stacking voids \cite{Hamaus_2015,Hamaus_2016, Cai_2016} -- these are no longer spherical in redshift space and may present a way of measuring the asymmetry along the line of sight and therefore give a new way to measure directly the relativistic effects in LSS. Future work could extend this analysis to higher-order statistics. Although relativistic corrections to the bispectrum and trispectrum have been derived, it is important to be able to investigate the specific distortions we have computed come into play within those higher order statistics, and how they might be detected observationally. In particular, it would be interesting to visualise these effects with realistic survey parameters, since magnification and evolution biases can alter the amplitude and even change the sign of some contributions at both first and second order. Moreover, a detailed comparison with numerical simulations that have these relativistic effects \cite{Adamek_2016} would help refine the theoretical predictions. 

\section*{Acknowledgements} We thank Yan-Chuan Cai and Chris Addis for helpful discussions and Roy Maartens for extensive discussions and valuable comments on the plots.
\bibliographystyle{JHEP}
\bibliography{ref}



\end{document}